\documentclass[useAMS,usenatbib]{mn2e}
\pdfoutput=1
\usepackage{enumitem}
\usepackage{natbib}
\usepackage{graphicx}
\usepackage{amssymb}
\usepackage{lineno}
\usepackage{amsmath}
\begin{document}
%%%%%%%%%%%%%%%%%%%%%%%%%%%%%%%%%%%%%%%
\newcommand{\MSun}{{M_\odot}}
\newcommand{\LSun}{{L_\odot}}
\newcommand{\Rstar}{{R_\star}}
\newcommand{\calE}{{\cal{E}}}
\newcommand{\calM}{{\cal{M}}}
\newcommand{\calV}{{\cal{V}}}
\newcommand{\calO}{{\cal{O}}}
\newcommand{\calH}{{\cal{H}}}
\newcommand{\calD}{{\cal{D}}}
\newcommand{\calB}{{\cal{B}}}
\newcommand{\calK}{{\cal{K}}}
\newcommand{\labeln}[1]{\label{#1}}
\newcommand{\Lsolar}{L$_{\odot}$}
\newcommand{\farcmin}{\hbox{$.\mkern-4mu^\prime$}}
\newcommand{\farcsec}{\hbox{$.\!\!^{\prime\prime}$}}
\newcommand{\kms}{\rm km\,s^{-1}}
\newcommand{\cc}{\rm cm^{-3}}
\newcommand{\Alfven}{$\rm Alfv\acute{e}n$}
\newcommand{\Vap}{V^\mathrm{P}_\mathrm{A}}
\newcommand{\Vat}{V^\mathrm{T}_\mathrm{A}}
\newcommand{\D}{\partial}
\newcommand{\DD}{\frac}
\newcommand{\TAW}{\tiny{\rm TAW}}
\newcommand{\mm }{\mathrm}
\newcommand{\Bp }{B_\mathrm{p}}
\newcommand{\Bpr }{B_\mathrm{r}}
\newcommand{\Bpz }{B_\mathrm{\theta}}
\newcommand{\Bt }{B_\mathrm{T}}
\newcommand{\Vp }{V_\mathrm{p}}
\newcommand{\Vpr }{V_\mathrm{r}}
\newcommand{\Vpz }{V_\mathrm{\theta}}
\newcommand{\Vt }{V_\mathrm{\varphi}}
\newcommand{\Ti }{T_\mathrm{i}}
\newcommand{\Te }{T_\mathrm{e}}
\newcommand{\rtr }{r_\mathrm{tr}}
\newcommand{\rbl }{r_\mathrm{BL}}
\newcommand{\rtrun }{r_\mathrm{trun}}
\newcommand{\thet }{\theta}
\newcommand{\thetd }{\theta_\mathrm{d}}
\newcommand{\thd }{\theta_d}
\newcommand{\thw }{\theta_W}
\newcommand{\beq}{\begin{equation}}
\newcommand{\eeq}{\end{equation}}
\newcommand{\ben}{\begin{enumerate}}
\newcommand{\een}{\end{enumerate}}
\newcommand{\bit}{\begin{itemize}}
\newcommand{\eit}{\end{itemize}}
\newcommand{\barr}{\begin{array}}
\newcommand{\earr}{\end{array}}
\newcommand{\bc}{\begin{center}}
\newcommand{\ec}{\end{center}}
\newcommand{\DroII}{\overline{\overline{\rm D}}}
\newcommand{\DroI}{{\overline{\rm D}}}
\newcommand{\eps}{\epsilon}
\newcommand{\veps}{\varepsilon}
\newcommand{\vepsdi}{{\cal E}^\mathrm{d}_\mathrm{i}}
\newcommand{\vepsde}{{\cal E}^\mathrm{d}_\mathrm{e}}
\newcommand{\lraS}{\longmapsto}
\newcommand{\lra}{\longrightarrow}
\newcommand{\LRA}{\Longrightarrow}
\newcommand{\Equival}{\Longleftrightarrow}
\newcommand{\DRA}{\Downarrow}
\newcommand{\LLRA}{\Longleftrightarrow}
\newcommand{\diver}{\mbox{\,div}}
\newcommand{\grad}{\mbox{\,grad}}
\newcommand{\cd}{\!\cdot\!}
\newcommand{\Msun}{{\,{\cal M}_{\odot}}}
\newcommand{\Mstar}{{\,{\cal M}_{\star}}}
\newcommand{\Mdot}{{\,\dot{\cal M}}}
\newcommand{\ds}{ds}
\newcommand{\dt}{dt}
\newcommand{\dx}{dx}
\newcommand{\dr}{dr}
\newcommand{\dth}{d\theta}
\newcommand{\dphi}{d\phi}

\newcommand{\pt}{\frac{\partial}{\partial t}}
\newcommand{\pk}{\frac{\partial}{\partial x^k}}
\newcommand{\pj}{\frac{\partial}{\partial x^j}}
\newcommand{\pmu}{\frac{\partial}{\partial x^\mu}}
\newcommand{\pr}{\frac{\partial}{\partial r}}
\newcommand{\pth}{\frac{\partial}{\partial \theta}}
\newcommand{\pR}{\frac{\partial}{\partial R}}
\newcommand{\pZ}{\frac{\partial}{\partial Z}}
\newcommand{\pphi}{\frac{\partial}{\partial \phi}}

\newcommand{\vadve}{v^k-\frac{1}{\alpha}\beta^k}
\newcommand{\vadv}{v_{Adv}^k}
\newcommand{\dv}{\sqrt{-g}}
\newcommand{\fdv}{\frac{1}{\dv}}
\newcommand{\dvr}{{\tilde{\rho}}^2\sin\theta}
\newcommand{\dvt}{{\tilde{\rho}}\sin\theta}
\newcommand{\dvrss}{r^2\sin\theta}
\newcommand{\dvtss}{r\sin\theta}
\newcommand{\dd}{\sqrt{\gamma}}
\newcommand{\ddw}{\tilde{\rho}^2\sin\theta}
\newcommand{\mbh}{M_{BH}}
\newcommand{\dualf}{\!\!\!\!\left.\right.^\ast\!\! F}
\newcommand{\cdt}{\frac{1}{\dv}\pt}
\newcommand{\cdr}{\frac{1}{\dv}\pr}
\newcommand{\cdth}{\frac{1}{\dv}\pth}
\newcommand{\cdk}{\frac{1}{\dv}\pk}
\newcommand{\cdj}{\frac{1}{\dv}\pj}
\newcommand{\rad}{\;r\! a\! d\;}
\newcommand{\half}{\frac{1}{2}}
%%%%%%%%%%%%%%%%%%%%%%%%%%%%
%%%%    Here

                  %\include{GlitchesPulsars_Hujeirat}
                  
\title[
Black holes, the Big Bang and the habitable universe: Are they really compatible?]
{Black holes, the Big Bang and the habitable universe: Are they really compatible? {}}
{}
\author[Hujeirat,  A.A.]
       {Hujeirat, A.A. \thanks{E-mail:AHujeirat@uni-hd.de} \\
\\
%  \footnotemark[1]\thanks{ }\\
%$^{1}$
IWR, Universit\"at Heidelberg, 69120 Heidelberg, Germany \\
%$^{2}$
}
\date{Accepted  ...}

\pagerange{\pageref{firstpage}--\pageref{lastpage}} \pubyear{2002}

\maketitle

\label{firstpage}

\begin{abstract}
 Astronomical observations have confirmed  the existence of BHs and the occurrence of the
 Big Bang event to beyond any reasonable doubt.  \\
 While quantum field theory and general theory of relativity predict the mass-spectrum of
 BHs to be unlimited, both theories agree that their creation is irreversible. \\
 In this article I argue that the recently-proposed SuSu-objects (: objects that are made of incompressible superconducting gluon-qurak superfluids), may not only entail the required properties to be excellent BH-candidates, but also encoding a hidden connection to dark matter and dark energy in cosmology. If such connection indeed exists, then the inevitable consequence  would be that our universe is infinite and subject to repeated Big Bang events of the second kind, which makes the habitability of the universe certain and
 our cosmic relevance insignificant and meaningless.

  \end{abstract}


\section{Introduction: the Big Bang and the escaped collapse into a black hole}
   Recent observations reveal that our observable universe harbors approximately
   $2\times 10^{18}$ galaxies, each contains   $\calO(10^8)$ stars on the average \citep{Conselice2016}.
   Assuming each star to consist of $10^{56}$ baryons, then the total energy of the luminous matter in our universe would be of order $10^{79}$ ergs. This amounts to $4.6\%$ of the total energy content of the universe, whereas the rest consists of non-baryonic matter and
   dark energy whose origin is a matter of debate.
   %%%%%%%%%%%%%%%%%%%%%%%%%%%%%%%%

\begin{figure}%[htb]
\centering {\hspace*{0.75cm}
\includegraphics*[angle=-0, width=7.15cm]{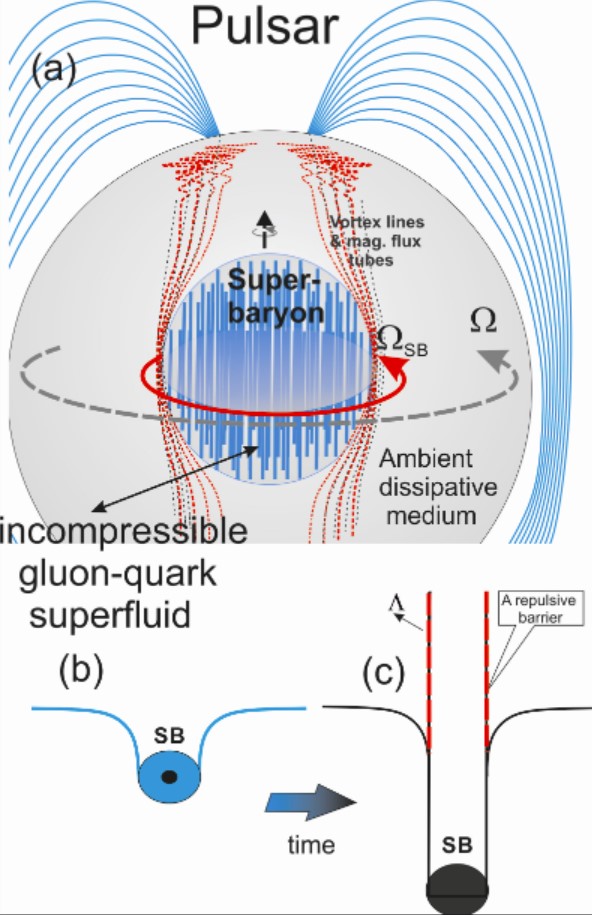}
}
\caption{\small Pulsars are born with relatively small embryonic super-baryons (SBs) at their centers (a). The cores of the SBs are made of incompressible superconducting gluon-quark superfluid (SuSu-Fluid), whereas the ambient medium is made of dissipative and conductive nuclear fluid and that rotates differentially i.e., $d\Omega/dr <0$.  The curved spacetime embedding the pulsars in combination with a scalar field at the background of supranuclear dense matter enforces the normal nuclear matter to undergo a phase transition into
SuSu-fluid, thereby enhancing its effective mass and increasing its red shift, up to the level that makes the object disappear from our observational windows. Similar to GQPs inside baryons in atomic nuclei,
SuSu-objects cannot exist in free space and so they shielded by a repulsive barrier that under normal conditions prohibit its merger with neighboring objects.
}  \label{XXX}
\end{figure}
%%%%%%%%%%%%%%%%%%%%%%%%%%%%%%%%%%%%%%
   %%%%%%%%%%%%%%%%%%%%%%%%%%%%%%%%
    If at the verge of the Big Bang the whole energy of the universe, namely $\calM_{universe} = \calO(10^{81})$ ergs, was compressed into a static sphere, whose radius coincides with
    the Schwarzschild one: $ R_S (\doteq 2G\calM_{universe}/c^2),$ then its average density would be  $<\rho>_i = a_0/\calM^2 = \calO(10^{-56}) $ g/cc.  However, the current density of the universe is approximately  $10^{27}$ times larger than $<\rho>_i$. Does it mean that we are living inside a BH? The fact that we are here means that the universe succeeded to escape its collapse during its birth. The mathematical reason therefor is that the spacetime at the verge of the Big Bang was neither curved nor static, but spatially flat and temporally rapidly changing \citep{Guth1997}.  \\
   Although the universe escaped its death at its birth, it was paradoxically unable to prohibit the formation of black holes during its subsequent evolutionary epochs. For example, shortly after the Big Bang explosion, it is believed that primordial density fluctuations collapsed to form primordial BHs, but due to their low masses, these objects must have  Hawking-evaporated by now \citep{Carr2004}. Later on, when the universe cooled down to several thousands degrees, the universe entered the phase of star formation. Sophisticated numerical simulations have shown that the first stars  formed during this epoch must have been massive. Due to their low metallicity, the central pressure in these massive stars  was relatively weak and fall to oppose the weight of the overlaying heavy shells,
   and so they collapsed under their own self-gravity to form the first generation of massive BHs. Subsequently, super-Eddington accretion in combination with and merger-events should have increased their masses to become the monstrous BHs lurking at the centers of almost all known galaxies in the observable universe.\\
   It should be noted however that due to the authority of Einstein, physicists ignored BH-physics for about half a century, but was revived eight years after Einstein passed away, when the first powerful quasars were discovered. \\
   Another strong evidence that favors the existence of BHs was announced in 2002 \citep{ESO2002}, when a German-team of ESO using approximately 16 years of observations of the stars in the vicinity of the Galactic center confirmed with unprecedented accuracy that the center must be occupied indeed by a monstrous black hole. \\

   However, once BHs are formed, the process is irreversible as their Hawking's evaporation time, is not only much longer than the age of the universe, but also,
   following \cite{Belinski1995},  neither particle creation nor emission of Hawking radiation at their event horizon are likely to occur. As noted by Hawking (2004): a distant observer cannot distinguish between real and virtual BHs and that quantum fluctuations would not allow the formation of true event horizons. \\
   
   I should note here that, if  BHs do radiate as black body, then their entropy must be large and  their temperature must be inversely proportional to their masses, i.e. $T_{BH} \propto 1/\calM_{BH}.$ This implies that the
   heat capacity must be negative: a result that must be ruled out under normal astrophysical conditions. Moreover, following Bekenstein (1974), BHs
   should have at least trillion times more entropy than what their progenitors originally
   had. However, as $T_{BH}> 0$, then the enormous entropy enhancement should be reflected
   as a dramatic increase of the effective mass-energy of the BH, which clearly counters observations.\\
   As a consequence, the so-called information paradox cannot be decoded and therefore is irrelevant for distant observers, as the time required to recover the information associated with particles crossing the event horizon is much longer than the age of any possible universe.
   
    Today there are many sophisticated observational methods that can be used to measure the  masses of BHs and other related properties precisely. Nonetheless, mathematically,
    all these methods will continue to be necessary for proving the existence of BHs, but by no way are sufficient.
    To clarify the argument: let a distant observer send a sophisticated satellite to measure the energetics of particles at the vicinity of the event horizon. No matter how long
    the observer will wait, he will be more and more sure that there is a black monster hidden behind the horizon, though its nature will continue to be undetectable. \\

    \section{Could SuSu-Objects resolve the BH-crises?}

   One possible resolution to this BH- crises is to conjecture that \textbf{natural physical objects, such as massive stars, cannot be transformed into singular physical entities, simply because the fundamental constants
    characterizing the physics of the universe we live in do not allow such eternal
    self-destructive events to occur.\\
     If such events did not occur at the verge of the Big Bang, why should  they occur during the succeeding less energetic epochs?  }
    In fact the cores of massive stars ought to cool relatively fast compared to their 
    low-mass counterparts and, and if they are sufficiently compact, they would end as BH-like objects. The process here evolves from inside-to-outside and relatively slowly compared to the dynamical time scale of the object.  In this case the formation of prompt   infinite barrier at the  horizon that separates the two Dirac seas can be completely avoided, thereby enabling observers to simultaneously collect information both from inside and outside the object.\\

    Indeed, ultra-compact objects (UCOs) such as pulsars, magnetars and neutron stars are the outcome of collapsed massive stars.
    These objects are extraordinary compact, $\alpha_S = R_S/R_\star \geq 1/2,$ and their mass-range is relatively narrow:  $1.3~\calM_\odot \leq \calM_{UCO} \leq 2.2~\calM_\odot.$
    By means of numerical simulation and using sophisticated equation of states, it was shown that:
    \begin{enumerate}[leftmargin=*,labelindent=1pt,label=\bfseries \arabic*.]
    %\begin{itemize}
      \item The densities at the centers of UCOs are much larger than the nuclear density, i.e. $\rho \geq \rho_0,$ though the physical laws governing the matter in this density-regime are neither clear nor verifiable.\\
          
      \item The nuclear fluid at the center is weakly incompressible, superconducting and
               in a superfluid phase.
    \een
    In fact recent experimental studies appear to indicate
    that gluon-quark plasmas, that make the cores of nucleons, is nearly perfect as the
    data seem to favor a much smaller viscosity over entropy ratio \citep{Romatschke2007}. In the case of UCOs, the spacetime embedding these objects would further enhance the compression of neutrons at their centers up to the level that new channels must be  created, through which the residual of the nuclear force is communicated efficiently. This in turn would enhance the effective energy density of the gluon-cloud,   rendering merger of neutrons to form super-baryon (SB) possible. The energy enhancement
    due to the creation of new communication channels can be viewed  as energy injection by a scalar field $\phi$ at the background, which
    becomes effective, whenever the  critical supranuclear density $\rho_{cr}$ is surpassed.
    The process has a run away character: the more energy is created inside the SB, the more curved will be the spacetime and therefore the more redshifted will be the UCO.\\
    Similar to protons whose lifetime is beyond $10^{30}$ yrs, it is reasonable to conjecture that at zero-temperature, there is just one single minimum energy state, $\Omega_S$, in which the gluons and quarks inside SBs can be organized, thereby giving rise to vanishing entropy: $dS = k_B ~log \Omega=0.$ Thus the pressure inside SBs cannot be local, as otherwise pressure waves would propagate randomly, collide with each other and lose energy; hence violating the minimum energy principle. When small external perturbation are set to hit the boundary of the SB, then the whole enclosed gluon-quark plasma is expected to react collectively to ensure
    global stability.   This requires that gluon-quark superfluids must have a uniform density and governed by a non-local negative pressure rather than by a local pressure: hence in a purely incompressible phase \citep{Hujeirat2016}.\\

    As I mention earlier, when two neutrons are brought together to merge, the scalar field  injects energy and enhances the effective mass of the resulting SB. Such an energy enhancement has been observed in the LHC-experiments during the years 2009-12, where
    formation of pentaquarks has been identified and whose rest energy was found to lay between 4.38-4.45 GeV \citep{LHCb2015}.\\
    Although the merger process inside UCOs is relatively slow, it would halt only, once the object has been entirely metamorphosed into a stellar-size SB-object. Given that
    the compactness parameter of UCOs $\alpha_S = R_S/R_\star \geq 1/2,$ then one may ask
    \textbf{whether the SB-object would continue to be luminous or would it collapse into a BH?}\\
    It should be emphasized here that the scalar field would need to at least
    double the mass of the UCO in order to ensue its collapse into a BH. However, stellar
     BHs with $\calM \leq 5~\Msun$ have never been observed.
    This implies that the dark energy enhancement by the scalar field is limited  and would not lead to a catastrophical self-collapse. Moreover,
    as UCOs with $\calM \geq 2.5~\MSun$ haven't been observed yet, we conclude that the compactness of stellar SBs must be $\alpha_S = 1  + \epsilon,$ where $\epsilon \ll 1$), i.e., they are deeply trapped in spacetime, highly redshifted and therefore would not appear in our observational windows.

  \section{SuSu-objects versus observations}

  SBs are macroscopic entity formed through fused baryons under the effects of strong gravitational fields in combination with a scalar field at the background of supranuclear dense matter.
  The dynamics and growth of SBs obey the laws of quantum dynamics, and in particular the
  Onsager-Feynman equation of superfluidity.
  %%%%%%%%%%%%%%%%%%%%%%%%%%%%%%%%%%%%%%
  In a previous article \citep{Hujeirat2017a}  I presented a scenario for explaining the origin and dynamics of the glitch-phenomena observed in pulsars and young neutron stars. Accordingly, when pulsars are born, embryonic SBs are simultaneously created  at their centers and start to gradually grow in mass with time. The cores of SBs are made of supranuclear superconducting gluon-quark superfluids having the constant density: $\rho_{SB} = 3\times \rho_0$ (see Fig. 1). Similar to the nuclear shell model, the mass and inertia of SBs are set to grow  with time following a well-defined  discrete quantum scheme to finally metamorphose the entire UCO into an invisible giant baryon.
The  ${}^4Helium-$superfluid inside a rotating container is strikingly similar
to the SB-UCO system. Let the initial rotation of the superfluid helium and the container
be equal. When  the container is subsequently spun-down, then the rotational frequency of the helium-superfluid would follow  a well-defined discrete quantum sequence $\{\Omega_{He}\},$ as dictated by the Onsager-Feynman equation.\\
The analogy to the SB-pulsars system is obvious: The SB corresponds to ${}^4Helium,$
whereas ambient~dissipative~medium to the container (: the rotational frequencies of both are determined by external torques). The observed glitch events would correspond then to a transition from one energy level to the next, or equivalently, from the spin frequency 
$\Omega^{n}_{SB}$ of the SB to the next lower one $\Omega^{n+1}_{SB}$. Due to the supercoductivity  and superfluidity character of the SB, the transition is associated with the ejection of a certain number of vortex lines into the ambient dissipative medium, which in turn absorbs and diffuses the vortices, thereby ensuing a prompt spin up of the crust, which is fully coupled to the ambient medium (Fig. 1). These prompt transitions
 are provoked mainly by the two mechanisms: 1) The continuous decrease of the rotational frequency of the ambient medium by magnetic torque of the pulsar and 2) Through the energy enhancement by the scalar field, which indirectly leads to a more compression and therefore
 to enhanced merger of neutrons.

\section{SuSu-objects and their universal impact}
  If SuSu-objects indeed exist in nature, then their impact on our understanding of the physics and evolution of our cosmos is far reaching. In the following I list a few of these consequences:
  %\begin{description}
   \begin{enumerate}[leftmargin=*,labelindent=1pt,label=\bfseries \arabic*.]
    %\item[1-]
    \item The density in our universe is upper-bounded by the universal
              critical density $\rho_{cr}\approx 3\times \rho_0.$\\
              Recalling that the lifetime of protons is much longer than the age of the universe, it reasonable to conclude that the quantum energy state of the gluon-quark cloud inside protons at zero-temperature should be the lowest possible energy state with vanishing entropy. The strong force
              between  quarks is communicated with maximum possible speed, yielding
              an EOS that converges towards $ P \rightarrow \calE = a~n^2.$  This corresponds to incompressible state of matter, as otherwise the causality principle would be violated.\\
              When neutrons at the center of UCOs are brought  to merge together,
              new communication channels are required for efficiently transmitting the strong force between them and ensure long-term stability.
              This however would enhance the total effective mass of the super-baryon, whereas the number density of quarks, whose contribution to the mass of the baryon is minor, will hardly change.  \\
              %%%%%%%%%%%%%%%%%%%%%
              Indeed, our theoretical studies have shown that the chemical potential at the centers of UCOs cannot grow indefinitely, but must converge to a constant value that corresponds to the critical density $\rho_{cr}=3\times \rho_0.$ The Gibbs function here attains a global minimum, and, in combination with the scalar field at the background, enables the normal compressible nuclear fluid to undergo a transition into incompressible quark-gluon superfluid phase.\\
              To summarize: the compression of nuclear fluids at the centers of UCOs is upper-limited as when  the density reaches $\rho_{cr}$,
              the fluid becomes purely incompressible.\\

    %\item[2-]
    \item  \textbf{BHs with true horizons do not need to exist.}\\
              When massive stars collapse to form UCOs, such as pulsars and neutron stars, the central density may surpass the nuclear density and reach the universal critical density $\rho_{cr} = 3 \times \rho_0,$ at which the nuclear fluid becomes purely incompressible. Thus, as long as the propagational speeds of external perturbations are superluminal, the gluon-quark cloud inside SBs  will have ample time to react comfortably and therefore avoiding their self-collapse.\\
              To clarify the point: consider a distant observer who is monitoring an ongoing collapse of a massive star. Having formed a SuSu-core of mass
              $\calM = \DD{1}{2}\calM_\odot$, then the enclosed fluid has roughly $2\times 10^{-5}$ s to react, which is equal to the light crossing time. However, the core wouldn't collapse into BH, unless the overlying shells of matter 
              collapse and hit the core of the SB at the rate of $\dot{\calM} \geq c^3/2G \approx 10^{5}~\calM_\odot$ per second, which is unrealistic even under extreme astrophysical conditions.
              
               \textbf{I should mention here that direct observations cannot confirm the existence of BHs, but that some sort of "black monsters" are hidden behind certain radii that are "roughly" close, but not equal to their corresponding true horizons.} Due to the incompressibility character 
               of gluon-quark-superfluids and their maximal compactness, SuSu-objects can neither exist in free space nor can be observed \citep{Witten1984}: hence making them to excellent BH-candidates.\\
%%%%%%%%%%%%%%%%%%%%%%%%%%%%%%%%%%%%%%

\begin{figure}%[htb]
\centering {\hspace*{0.75cm}
\includegraphics*[angle=-0, width=7.5cm]{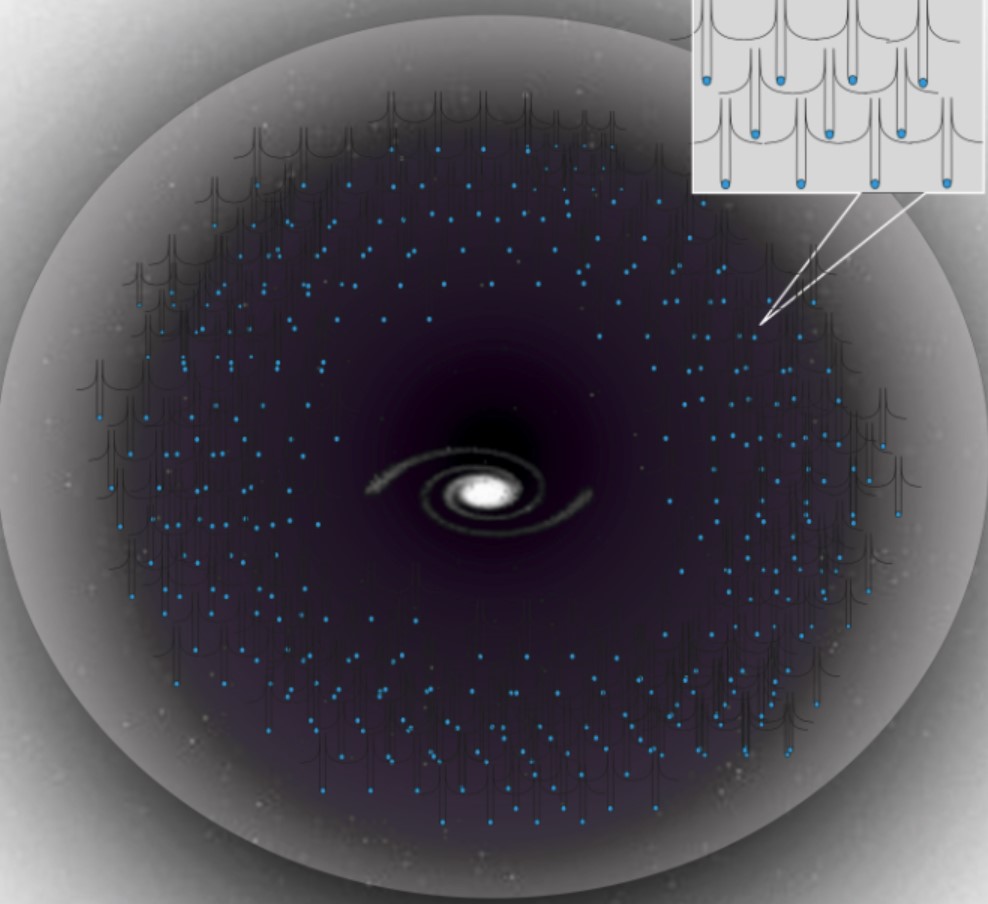}
}
\caption{\small   Dark matter halos consist of
weakly interacting SuSu-objects that have conglomerated over several big bang epochs
into clusters, which are observed to embed  the galaxies in the observable universe.
The SuSu-objects are maximally compact and therefore deeply trapped in spacetime.
This make halos look like forests made of billions of needle-like curved spacetime.
}  \label{XXX}
\end{figure}
%%%%%%%%%%%%%%%%%%%%%%%%%%%%%%%%%%%%%%
    %\item[3-]
    \item \textbf{Do SuSu-objects have the potential to resolve the mystery of the dark matter
              and dark energy in cosmology?}\\
              Almost all galaxies in the observable universe are considered to be embedded in dark matter (DM) halos \citep{Yoshida2000}, which regulate the evolution and dynamics of galaxies. Other than their gravitational interaction with the baryonic matter, they don't show any observational signatures that could unveil their true nature, hence leaving both theoreticians and observers alike to speculate about their origin and nature.  \\
              
              The main properties of dark matter read as follows:\\
              
              \begin{itemize}
                \item Dark matter is non-luminous, weakly-interacting particles/objects
                \item They interact with baryonic matter solely via gravitation
                \item The matter comprising DM is electromagnetically neutral and does not contain anti-particles
                \item The mass of DM in the universe is almost five time as large as the baryonic matter.
              \end{itemize}

              In fact SuSu-objects fulfill all these properties nicely: a stellar-size SB object would be deeply trapped in spacetime and its corresponding redshift would be so large that they look completely black (maximum compactness). As in atomic nuclei, where each nucleon is shielded by a repulsive barrier that forbid its fusion with its neighbors, we anticipate SuSu-objects to behave similarly and to conglomerate uniformly into clusters, which can be viewed as forests made of billions of needle-like curved spacetime embedded in the globally curved spacetime (Fig. 2).

              Thus, unless the spacetime embedding SuSu-objects and the repulsive barrier shielding them have limited lifetime, these objects would have enough time
              to gravitationally control the dynamics of the enclosed galaxies.\\

    %\item[3-]
    \item Our universe is infinite and subject to repeated  big bang events of the second kind.\\
             In fact UCOs are expected to cool and to metamorphose into SuSu-objects on the time scale of several million years up to one Gyr after their births. However, their conglomeration into halos would last longer
             than the age of the embedded galaxy, or even than the age of the universe. This would simply imply that dark matter halos may contain relics of SuSu-objects that  have formed  during successive big bang epochs.  \\

\een
  %\end{description}

Unlike the binding energy of nucleons inside atomic nuclei, which peaks around iron-56, the deconfinement energy of gluon-quark superfluids (henceforth GQ-superfluids) inside SBs would continue to grow with the baryon number, attaining maximum at the event horizon plus epsilon.

Similar to GQ-plasmas (GQP) inside individual baryons, the ocean of the GQ-superfluid inside SuSu-objects most likely would be shielded from the outside world by a quantum repulsive membrane. We conjecture that this membrane would be sufficiently strong to inhibit quantum tunneling of particles from both inside and outside of the barrier, as otherwise DM halos would be electromagnetically active.  
In this case there must be a length scales $\Lambda_{rm}$, such that, when the separation length, $d,$ between two arbitrary SuSu-objects becomes smaller or comparable to $\Lambda_{rm},$ then, similar to nucleons in atomic nuclei, the repulsive forces will dominate and prohibit their fusion/merger (see Fig. 1/c).\\

Assume the dark matter halo embedding the Milky Way  to consist of SuSu-objects that have conglomerated into a cluster over several big bang epochs. Because of unknown physical mechanism, the strongly curved spacetime embedding each object becomes dynamically unstable. The nature of SuSu-objects, in particular its zero-temperature superfluidity and zero-entropy, most likely would enable them to intercommunicate with each other via De Broglie waves, whose wave-lengths can be extra-ordinary large at very low temperatures ($\lambda_{DB} \propto 1/T$).
This instability could ensue collective collisions and mergers giving rise to powerful fireworks, through which approximately  $10^{66}$ ergs would be released instantly, hence
sufficiently  energetic to blow up and flatten the spacetime and set the enclosed matter into divergent motions.

 Such an
event  would correspond to a  big bang explosion of second kind.
 In computing the released energy, I assumed that the dark matter halo surrounding the Milky Way consists of approximately  $4\times 10^{12}\, \MSun$ of SuSu-objects.
 The dark energy stored in each object is exactly equal to its baryonic mass, as the compactness parameter of UCOs at their birth is assumed to be half. This
 yields a total dark energy of order $2\times 10^{12}\, \MSun\times c^2,$ which should be instantly released during the explosion. The superanuclear dense matter is opaque
  and therefore  the released energy would be thermalized, giving rise to the effective temperature: $ T = 0.949~GeV/k_B \approx 10^{13}$\,K, which is sufficient high to keep the soup of the gluon-quark particles hot for a while.
 
%%%%%%%%%%%%%%%%%%%%%%%%%%%%%%%%%%%%%%

\begin{figure}%[htb]
\centering {\hspace*{0.75cm}
\includegraphics*[angle=-0, width=7.75cm]{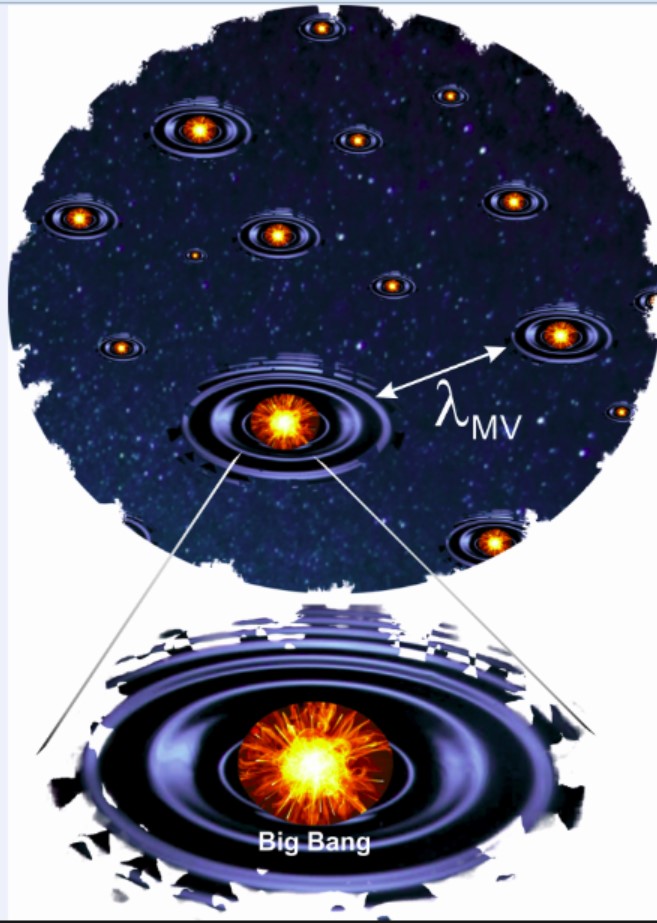}\\
}
\caption{\small The strongly curved spacetime embedding each SuSu-object in the halo most likely have limited lifetime and therefore they ought to decay. The decay time certainly is longer than 13.8 Gyrs, but most likely shorter than the decay time of protons $\tau_p$. Hence the proper distance  between two arbitrary big bang events in this infinite/multiverse universe  would be $\lambda_{MV}= c\,\tau_p.$ This in turn ensues collective collisions and mergers of the whole objects in the halo, igniting thereby a cosmic firework that we term here big bang of second kind. The released energy would be sufficient to inflate the spacetime and sets its contents into  divergent motions.
}  \label{XXX}
\end{figure}
%%%%%%%%%%%%%%%%%%%%%%%%%%%%%%%%%%%%%%
 In fact the  evolutionary scenario presented here is in line with the cosmological evolution of  the chemical abundance of our universe. When the billions of SuSu-objects undergo an instantaneous merger, the released dark energy  will enforce the
 super-baryons to decay back into their normal-size baryons, though extra-ordinary heated by the thermalized dark energy. The prompt expansion of the local spacetime would cause the soup of GQP to rapidly cool down, recombine and form the first light nuclei and evolve in the manner that predicated by the classical Big Bang theory.

I should stress here that the consistency of an ever expanding, isotropic and homogeneous universe and its origin from just one single giant Big Bang event is not at odd with here-presented scenario. However, the classical picture it should be modified to include
the new mechanism underlying the Big Bang explosion, that our universe is infinite and is subject to repeated big bang events of the second kind. 
Indeed, the possibility that certain galaxies and therefore their DM-halos could be older than the age of the universe should not be ruled observationally. \\

\bit
 \item[5-] The consensus that the fundamental constants in our local universe are indeed spatially and temporarily constant should apply still to our infinite universe. \\
            Let the dark matter halos embedding primordial galaxies contain relics of SuSu-objects. Since these objects have evolved from normal baryon matter inside the galaxies, we expect them to obey to the same laws of physics that govern our galaxy.  Indeed, if annihilation of SuSu-objects is the mechanism that ignite big bang explosions, then is unlikely that the objects formed in the succeeding epochs would evolve according to different fundamental constants and produce exotic objects.\\

    \item[6-] The fundamental constants of our infinite universe together with the repeated big bang events of second kind (i.e. the universe is capable of withstanding
        unlimited number of big bang events at the same time with respect to a distant observer) would certainly have sufficient habitable planets for
        developing biological organisms and ensuring their reproduction.\\

        In fact most of the stars in the observable universe are believed to be surrounded by planets. Based on recent observations, it was found that 21 out of the 5000 exoplanet-candidates in principle could have Earth-like size with environmental conditions that could be suited for developing microbial organisms \citep{Nasa2017}.\\

Recalling that there are at least additional $10^{18}$ galaxies in the observable universe, then the probability of having habitable planets with developed biological organisms cannot be diminishingly small. However, this probability would converge even to one, if the observable universe is just a negligible small portion of an infinitely large, isotropic and homogeneous universe.\\

\eit
\section{Conclusions}
The proposal that UCOs, such as pulsars, magnetars and neutron stars
metamorphose into extraordinary massive super-baryons, whose interiors are made of incompressible superconducting gluon-quarks superfluids and whose radii are equal to the
corresponding event horizons plus epsilon would have far reaching consequences on our understanding of particle physics, the physics of the Big Bang and the cosmology of our universe.\\
Among the important consequences are:1) there exist a maximum critical energy density that cannot be surpassed by any possible astrophysical event in the universe. 2) Massive stars
may undergo dramatic changes in their evolution, though they remain real physical objects and would not collapse directly into BHs, whose physical entities cannot be decoded by distant
observers. 3) Dark matter may be composed of SuSu-objects that have conglomerated into halos
over several big bang epochs. 4) The universe is infinite and subject to indefinite number of repeated big bang events of the second kind. 5) The fundamental constants in this infinite universe are indeed specially and temporally constant, so that the probability of having habitable planets in our infinitely large universe converges almost to one. This however emphasizes our cosmic insignificance and diminishes the reasoning for our existence.\\
An interesting consequence of this scenario is that, while each universe has a non-diminishing probability to develop  biological organisms within, say the lifetime of protons,  it is certain that these will be entirely erased through the succeeding big bang explosions.

Finally, let me note that the present scenario agrees well with the
intuitive prediction of Albert Einstein that the universe static and infinite and that
nature does not allow its self-destruction such as forming BHs.


\begin{thebibliography}{99}
\bibitem[Bekenstein (1972)]{Bekenstein1972} Bekenstein, J.D., Phys. Rev. D, 7, 2333, 1974
\bibitem[Belinski (1995)]{Belinski1995} Belinski, V.A., Physical Lett., A209, 13, 1995
\bibitem[Carr (2004)]{Carr2004} Carr, B.J., in "22nd Texas Symposium on Relativistic
          Astrophysics at Stanford University", 1, 2004
\bibitem[Conselice et al. (2016)]{Conselice2016} Conselice, C.J., Wilkinson, A., et al., ApJ, 830, 83, 2016
\bibitem[Guth (1997)]{Guth1997} Guth, A., Ed. in "Theory of Cosmic Origins", Basic Books. 233, 1997
\bibitem[Hawking (2004)]{Hawking2014} Hawking, S., Note in: 17th Int. Conf. of GR and Gravit., 2004
\bibitem[Hujeirat (2012)]{Hujeirat2012} Hujeirat, A.A., MNRAS, 423.2893, 2012
\bibitem[Hujeirat (2016)]{Hujeirat2016}  Hujeirat, A.A., arXiv: astro-ph 1604.07882, 2016
\bibitem[Hujeirat (2017a)]{Hujeirat2017a}  Hujeirat, A.A., arXiv: astro-ph 1705.06608, 2017 (a)
\bibitem[LHCb Collaboration  (2015)]{LHCb2015} LHCb Collaboration, Phys. Rev. Lett., vol. 115, 2015
\bibitem[NasaExoplanet-Archive (2017)]{Nasa2017} NasaExoplanert-Archive, https://exoplanetarchive.ipac.caltech.edu, 2017
\bibitem[Romatschke \& Romatschke (2007)]{Romatschke2007} Romatschke, P. and Romatschke, U., Phys. Rev. Lett. 2007
\bibitem[Schoedel et al. (2002)]{ESO2002}  R. Schoedel, R., Ott,T., Genzel, R., et al., Nature, 419, 694, 2002
\bibitem[Witten (1984)]{Witten1984} Witten, E., Physical Review D, Volu. 30, Issue 2,  1984
 \bibitem[Yoshida et al. (2000)]{Yoshida2000} Yoshida, N., Springel, V., White, S.D.M., et al.,
  ApJ, 544, L87, 2000
 \end{thebibliography}
\end{document}